\documentclass[twocolumn,10pt]{IEEEtran}

\usepackage{amsmath}%\documentclass[12pt, draftclsnofoot, onecolumn]{IEEEtran}
\usepackage{amsfonts}
\usepackage{epsfig}
\usepackage{amssymb}
\usepackage{cite}
\usepackage{color,soul}
\usepackage{subfigure}
\usepackage{multirow}
\usepackage{rotating}
\usepackage{graphicx}
\usepackage{tabularx}
\usepackage{array}
\usepackage{graphicx,dblfloatfix}
\usepackage{blindtext}
\usepackage{ragged2e}
\usepackage{xcolor,colortbl}
\definecolor{Gray}{gray}{0.85}
\usepackage{amsthm,amssymb,amsmath,bm}
\usepackage{fancyhdr}
\usepackage[subfigure]{tocloft}
\usepackage[font={small}]{caption}
\usepackage{algorithmic}
\usepackage[Algorithm]{algorithm}

\begin{document}
 \title{Bit Error Probability Instead of Secrecy Rate Criterion to Enhance Performance for Secure Wireless Communication Systems} 
\author{\IEEEauthorblockN{Javad Taghipour, Paeiz Azmi,  Nader Mokari,  Moslem Forouzesh, and Hossein Pishro-Nik }
\textsuperscript{}\thanks{\noindent\textsuperscript{} Javad Taghipour is with the Department of Electrical and Computer Engineering, Tarbiat Modares University, Tehran, Iran.
	
	Paeiz Azmi is with the Department of Electrical and Computer Engineering, Tarbiat Modares University, Tehran, Iran..
	
	Nader Mokari is with the Department of Electrical and Computer Engineering, Tarbiat Modares University, Tehran, Iran.
	
	Moslem Forouzesh is with the Department of Electrical and Computer Engineering, Tarbiat Modares University, Tehran, Iran.

	H. Pishro-Nik is with the Electrical and Computer Engineering Department, University  of Massachusetts, Amherst, MA, 01003, USA.
}}
 \maketitle
  \begin{abstract}
In this paper, we propose a new practical power allocation technique based on bit error probability (BEP) for physical layer security systems. It is shown that the secrecy rate that is the most commonly used in physical layer security systems, cannot be a suitable criterion lonely. Large positive values are suitable for the secrecy rate in physical layer security, but it does not consider the performance of the legitimate and adversary users. In this paper, we consider and analyze BEP for physical layer security systems because based on it, the performance of the legitimate and adversary users are guaranteed and it is needed to use lower power. BEP is calculated for the legitimate and adversary users and it is shown that BEP can be better criterion for performance evaluation of the physical layer security systems. Based on BEP, the optimum transmit power is obtained and a new definition for outage probability is proposed and obtained theoretically. Also, the proposed approach is applied for adversary users with unknown mode and the cooperative adversary users. Simulation results show that the proposed method needs more than 5dB lower power for different scenarios. 
  	
 \end{abstract}
 
 \section{Introduction}\label{Introduction}
 
 Physical layer security that is introduced by Wyner [1] is one of the hot topics in secure wireless communication networks. By employing physical layer security, there is no need to use encryption, key management and the other complicated scenarios to apply security in the network. Therefore, it can be used in low power and low required computation networks like internet of things (IoT) or wireless sensor networks in 5G. In physical layer security, secrecy rate is the most important criterion that is used for performance analysis or resource allocation procedures. The secrecy rate is defined as the difference of the capacities of the legitimate user and adversary user [2].
 In recent years, a lot of scenarios and resource allocation methods are proposed in physical layer security systems. The most important purpose of these methods is maximizing the secrecy rate. To maximize the secrecy rate and apply physical layer security in these networks, transmission of artificial noise [3]-[6] and multiple antennas systems [7]-[9] are commonly used methods.
 As we know, there are two conventional categories of jamming signal that are considered to enhance physical layer security in wireless networks:
 \begin{itemize}
 	\item 	Friendly jamming: In this type of cooperative jamming, the jamming signal is trained to the legitimate receiver before network setup [10]-[12].
 	\item	Gaussian noise jamming: In this type of cooperative jamming, the jamming signal is unfamiliar at the legitimate receiver. Note that employing each of the friendly jamming or Gaussian noise jamming methods enhance security [13]-[14].
 \end{itemize}
 It is clear that friendly jamming causes to increase secrecy performance with respect to Gaussian noise jamming. Since the legitimate receiver is able to eliminate jamming signal from the received signal, hence the secrecy rate increases. It should be noted that this enhancement in performance imposes higher implementation complexity compared with Gaussian noise jamming. Moreover, jamming transmission is performed by the following methods, 1) Source- based jamming [15], [16], in this method source combines information signal and jamming signal, then transmits them together. 2) External nodes-based jamming: in this method a relay [17]–[18] or an external jammer [19] transmits jamming signal. 3) Legitimate destination-based cooperative jamming: in this method legitimate destination transmits jamming signal to defraud adversary users [20]–[22].
 
 Combining of transmit jamming signal or artificial noise and using multiple antenna attract many attentions. For example, transmission of spatially artificial noise signal by using multiple antenna and beamforming is proposed in [23].
 
 It is necessary to note that physical layer security always faces with some challenges like channel state information (CSI) and information about the modes or locations of the adversary users. To tackle these challenges, the authors in [24] investigate physical layer security for imperfect channel state information. Moreover, unknown modes of the adversary users and unknown location of the adversary for the legitimate users are studied in [25] and [26], respectively.
 
 In all the mentioned works, the evaluation criterion is the secrecy rate. The secrecy rate should be a positive value and it is suitable to be a large positive value. In physical layer security, the capacity of the adversary users is ignored, and the difference of the capacities is important. In some applications and services like voice, it can be more effective and the adversary users can detect the information signal with the desired performance. For a positive value of the secrecy rate, the adversary users can receive the data information with the desired bit error probability (BEP). The BEP can be another criterion to prevent data detection of the adversary users. In this paper, we consider BEP as a new criterion for power allocation of the physical layer security systems.
 
In [27]-[29], the low density parity check (LDPC) channel coding scheme is used to reach the desired bit error rate (BER) values for the adversary and legitimate users. By using the punctured binary LDPC code, the SNR gap in a secure system is decreased, and the adversary user cannot detect the data information [27]. These works are related to weak secrecy or strong secrecy, and they do not consider resource allocation in secure network. Also, in these methods, the adversary users do not know the parity check matrix, because this matrix is made by secret keys and just the legitimate users know about that. This assumption applies a limitation for the adversary users and it is in conflict with the physical layer security methods. 

In many applications in 5G communication systems like IoT, device to device (D2D) communication, wireless sensor network and etc., users cannot utilize high complexity and high power consumption transmitters and receivers. Therefore, in some applications like sensor networks, they cannot use the complicated channel coding schemes like LDPC. Also, if the adversary users know about the channel decoding mechanism, they can use the benefit of the channel coding for themselves. 
 
  In this paper, to avoid detecting of the data information by the adversary users, based on the type of services, we define a threshold value for BEP. In the proposed power allocation problem, the BEP of the adversary users should be more than the threshold value to prevent data detection by the adversary users with the desired performance. Also BEP of the legitimate users should be less than a threshold value to detect data information in legitimate receivers with the desired performance. The modulation type that is used in this paper, is binary phase shift keying (BPSK), and it is expanded to the other modulation types. Based on BEP and threshold values, the optimum power value is obtained. 
 
 The outage probability is one of the most important issues in wireless communication networks. In this paper, a new definition for the outage probability is proposed based on BEP. The outage probability that is proposed in this paper is the probability of the loosing of legitimate receiver performance or probability of reaching the adversary receiver to a desired performance. The outage probability is obtained analytically and it is approved by simulation results.
 
 In most of the works in physical layer security systems, the adversary users are assumed active or passive. In this paper, we assume that there is no information about active or passive mode of the adversary users and it is an unknown information. Also, for the multi adversary users’ case, it is assumed that they can have cooperation with each other to enhance their performance. 
 
 Similar to the commonly used methods for the physical layer security systems, in this paper, the multiple antenna system and artificial noise are used to decrease the outage probability of the network. Also, a simple channel coding scheme is proposed to enhance the performance of the legitimate receiver without applying any limitation to the adversary receivers. By using the proposed channel coding scheme, the outage probability of the network is decreased.
 
 The summary of the contributions are as follows:
 \begin{itemize}
 	\item 	Introducing novel power allocation problem by using a new BEP based criterion. Based on this problem, performance of the legitimate user is guaranteed to be in the desired area and the performance of the adversary users is guaranteed not to be in the desired area.
 	\item	Obtaining the optimum power value for the proposed optimization problem.
 	\item	Proposing a new definition for the outage probability. 
 	\item	Defining the power allocation problems and solve them for multi adversary users, unknown mode and cooperative adversary user’s scenarios.
 	\item	Using multiple antenna transmitter and channel coding to decrease the outage probability.
 	\item	Based on simulation results, it is shown that the proposed method guarantees the performance of all network users according to BEP criteria and it needs about 5dB lower transmit power.
 	
 \end{itemize}
 The rest of the paper is organized as follows. The secrecy rate and study motivation are presented in Section \ref{motivation}. The proposed BEP based power allocation, obtaining optimum transmit power and the new definition for the outage probability are presented in Section \ref{power allocation}. Improving the outage probability by using multiple antennas transmitter and channel coding is proposed in Section \ref{multiple antennas}. The performance evaluation and the simulation results are discussed in Section \ref{results}. Finally, concluding remarks are presented in Section \ref{Conclusion}. 
 
 \section{The secrecy rate and study motivation}\label{motivation}
 We consider a wireless communication network consisting of a transmitter, one legitimate and an eavesdropper or adversary user. The secrecy rate for this network can be obtained as follows:
 \begin{align} \label{RS}
 {R_s} = {\left[ {{{\log }_2}\left( {1 + SN{R_B}} \right) - {{\log }_2}\left( {1 + SN{R_E}} \right)} \right]^ + },
 \end{align}
 where $SN{R_B}$ and $SN{R_E}$  are the signal to noise power ratio (SNR) of the legitimate and adversary receivers  and ${\left[ x \right]^ + }$ denotes $\max \left\{ {x,0} \right\}$. SNR values in this system can be presented as follows:
 
  \begin{align}
&SN{R_B} = \frac{{\alpha _B^2{{\left| {{h_B}} \right|}^2}P}}{{{\sigma ^2}}}, \\&
SN{R_E} = \frac{{\alpha _E^2{{\left| {{h_E}} \right|}^2}P}}{{{\sigma ^2}}},
 \end{align}
 where $\left| . \right|$ is the absolute value,  $P$ is the transmit power, ${h_B}$
 is the channel coefficient of the legitimate user with path loss ${\alpha _B}$,  ${h_E}$ is the channel coefficient of the adversary user with path loss ${\alpha _E}$ and ${\sigma ^2}$ is the additive noise power in the legitimate receiver and adversary user. The channel coefficients are assumed to be independent identically distributed (i.i.d.) zero-mean circularly symmetric complex Gaussian random variables with unit variance.
 
 \textbf{Lemma 1.} . The secrecy rate that is defined in \eqref{RS}, is an increasing function for $SN{R_B} \ge SN{R_E}$.
 
 \textit{Proof}: Based on \eqref{RS}, the first order derivative function of the secrecy rate can be calculated as follows:
 \begin{align}
&\frac{{\partial {R_s}}}{{\partial P}} =\\& \left\{ \begin{array}{l}
\frac{1}{{\log 2}}\left( {\frac{{\alpha _B^2{{\left| {{h_B}} \right|}^2}}}{{{\sigma ^2} + \alpha _B^2{{\left| {{h_B}} \right|}^2}P}} - \frac{{\alpha _E^2{{\left| {{h_E}} \right|}^2}}}{{{\sigma ^2} + \alpha _E^2{{\left| {{h_E}} \right|}^2}P}}} \right),SN{R_B} \ge SN{R_E}\\
0, \hspace{5.2cm}SN{R_B} < SN{R_E}.
\end{array} \right.\nonumber
 \end{align}
 According to this equation, the secrecy rate obtained in (1), is an increasing function for $SN{R_B} \ge SN{R_E}$. If the channel power gain of the legitimate user is more than that of the adversary user, the first derivative function of the secrecy rate has a positive value and then the secrecy rate function will be increased by increasing the transmit power. $\blacksquare$
 
 In this case, because the secrecy rate increases by increasing the transmit power, the optimum value for the transmit power is its maximum value. The maximum power that the transmitter can transmit is the optimum value for the secrecy rate, and transmitter increases transmit power up to the maximum value.
 On the other side, the secrecy rate in (1) has a limited value and it is not increased to infinity by increasing the transmit power. The upper bound of the secrecy rate can be obtained as follows:
 
  \begin{align}
\mathop {\lim }\limits_{P \to \infty } {R_s} = \left\{ \begin{array}{l}
{\log _2}\left( {\frac{{\alpha _B^2{{\left| {{h_B}} \right|}^2}}}{{\alpha _E^2{{\left| {{h_E}} \right|}^2}}}} \right){,_{}}_{}SN{R_B} \ge SN{R_E}\\
0, \hspace{2.13cm}SN{R_B} < SN{R_E}.
\end{array} \right.
 \end{align}
 Therefore, the secrecy rate value for this system has a maximum value as follows:
 \begin{align}
 {R_{s,\max }} = {\log _2}\left( {\frac{{\alpha _B^2{{\left| {{h_B}} \right|}^2}}}{{\alpha _E^2{{\left| {{h_E}} \right|}^2}}}} \right).
 \end{align}
 The secrecy rate cannot increase to more than this value by increasing the transmit power. The secrecy rate is saturated for the large values of the transmit power. Therefore, it is not practical to increase the transmit power up to the large values.
 
  To avoid using maximum transmit power, the optimization problem with minimizing transmit power for a guaranteed secrecy rate can be used. This problem can be defined as follows:
 \begin{align}
 \begin{array}{l}
 \mathop {\min }\limits_P P\\
 s.t. \hspace{.2cm}{R_S} \ge {R_{min}},
 \end{array}
 \end{align}
 where ${R_{min}}$ is the minimum value for the secrecy rate that should be guaranteed for the network. For this problem, the optimum value of the transmit power is the minimum value of the transmit power that satisfies the constraint of the problem. For this problem, there is no need to increase the transmit power to the maximum value.
 
 The important issue is the work point or the selected value for the transmit power. If we use the maximum value or a large value for the transmit power, the secrecy rate is a desired value. But by using the maximum power value, the adversary user can receive signal with high SNR value, and he/she can detect data signal with the desired performance. In this paper, we consider this issue from the performance view of the legitimate receiver and the adversary user.
 
 Therefore, we can choose another criterion to obtain the transmit power value in the transmitter. One of the most important criteria for performance evaluation in digital communication systems is BEP. BEP can be used for performance evaluation of the legitimate receiver and the adversary user. 
 
 Based on the different services, the BEP can have different values in communication networks. For example, in long term evolution (LTE) systems, the maximum value of BEP for voice services is larger than that for signaling or synchronization [30]. 
 
 For the legitimate users and adversary users, by increasing the transmit power value, BEP for both of them is decreased. For example, if BPSK is used for a communication system, increasing the transmit power to the maximum value or a large value reduces the BEP for both legitimate receiver and the adversary user.
 
 In the mentioned system model, for a positive secrecy rate, BEP of an adversary user is more than that of a legitimate receiver. Although the BEP value for adversary user is more than that of legitimate receiver, this value can be acceptable for some services. Then the adversary user can receive and detect data information without any degradation in the performance. Therefore, the secrecy rate cannot be a sufficient criterion for these systems.
 BEP of BPSK signaling for a specific channel coefficient is obtained as follows [31]:
  \begin{align}
BEP = Q\left( {\sqrt {SNR} } \right),
 \end{align}
 where $Q\left( . \right)$ is the q function. Based on this equation, BEP is exponentially decreased by increasing the transmit power for both legitimate receiver and the adversary user. By increasing the transmit power to the maximum value, the BEP for the adversary user can be decreased to the desired value for a specific service. In this paper, we consider BEP of the users in power allocation to prevent this problem. 
 
 \section{Proposed BEP based power allocation }\label{power allocation}
 \subsection{BPSK signaling}
 In this section, the power allocation problem is defined based on BEP of the legitimate receiver and the adversary user. According to the service type, two threshold values are defined for the legitimate receiver and adversary user. When BEP of the legitimate receiver is less than a threshold value, he/she can receive information data without any degradation. If BEP of the adversary user is more than a threshold value, he/she cannot receive data information efficiently. 
 
 Based on these threshold values, the power allocation problem can be defined as follows:
 \begin{align}
\begin{array}{l}
\mathop {\min }\limits_P P\\
s.t. \,\, BE{P_B} \le {T_1},\\
\hspace{.52cm}BE{P_E} \ge {T_2},
\end{array}
 \end{align}
 where $BE{P_B}$ and $BE{P_E}$ are BEP of the legitimate user and adversary user respectively, $T_1$ is the maximum value for the $BE{P_B}$ and $T_2$ is the minimum value for  $BE{P_E}$. According to this problem, BEP for the legitimate user must be less than a threshold value $T_1$ to reach an efficient performance, and BEP for the adversary user must be more than a threshold value $T_2$, it causes that the adversary user cannot detect data information efficiently. 
 In the above problem, if BPSK is used for signaling, it can be rewritten as follows:
  \begin{align}
  \begin{array}{l}
  \mathop {\min }\limits_P P\\
  _{}s.t. \,\, Q\left( {\sqrt {\frac{{\alpha _B^2{{\left| {{h_B}} \right|}^2}P}}{{{\sigma ^2}}}} } \right) \le {T_1},\\\hspace{.52cm}
  Q\left( {\sqrt {\frac{{\alpha _E^2{{\left| {{h_E}} \right|}^2}P}}{{{\sigma ^2}}}} } \right) \ge {T_2}.
  \end{array}
  \end{align}
  Without loss of generality, in this paper we assume that the power and energy values are the same, and we use power value instead of energy value in BEP relations. To solve this problem, it can be rewritten as follows:
 \begin{align}
 \begin{array}{l}
 \mathop {\min }\limits_P P\\
 _{}s.t. \,\, P \ge \frac{{{\sigma ^2}}}{{\alpha _B^2{{\left| {{h_B}} \right|}^2}}}{\left( {{Q^{ - 1}}\left( {{T_1}} \right)} \right)^2},\\\hspace{.52cm}
 P \le \frac{{{\sigma ^2}}}{{\alpha _E^2{{\left| {{h_E}} \right|}^2}}}{\left( {{Q^{ - 1}}\left( {{T_2}} \right)} \right)^2},
 \end{array}
 \end{align}
 where ${Q^{ - 1}}\left( . \right)$ is the inverse q function. To solve this problem, it should be a feasible problem. According to the power constraints, the feasibility condition of this problem can be obtained as follows:
 \begin{align}
 \alpha _E^2{\left| {{h_E}} \right|^2}{\left( {{Q^{ - 1}}\left( {{T_1}} \right)} \right)^2} \le \alpha _B^2{\left| {{h_B}} \right|^2}{\left( {{Q^{ - 1}}\left( {{T_2}} \right)} \right)^2},
 \end{align}
 For a feasible problem, the optimum solution for the transmit power can be calculated as follows:
  \begin{align}
{P_{opt}} = \frac{{{\sigma ^2}}}{{\alpha _B^2{{\left| {{h_B}} \right|}^2}}}{\left( {{Q^{ - 1}}\left( {{T_1}} \right)} \right)^2}.
 \end{align}
 For this system, we propose a new definition for the outage probability as follows:
 \begin{align}
&p_{Outage} =\\& 1 - \Pr \left\{ {\alpha _E^2{{\left| {{h_E}} \right|}^2}{{\left( {{Q^{ - 1}}\left( {{T_1}} \right)} \right)}^2} \le \alpha _B^2{{\left| {{h_B}} \right|}^2}{{\left( {{Q^{ - 1}}\left( {{T_2}} \right)} \right)}^2}} \right\}, \nonumber
 \end{align}
 where  $\Pr \left\{ . \right\}$ is the probability value function. This outage probability shows the outage of the network from desired performance of the legitimate user and secure communication simultaneously. This outage can be because of unsecure communication or performance degradation of the legitimate user.
 
 Because the channel coefficients are i.i.d. zero-mean circularly symmetric complex Gaussian random variables with unit variance, channel power gains are exponential random variables. Therefore, the probability of outage can be obtained as follows:
  \begin{align}
{p_{Outage}} = \frac{{\alpha _E^2{{\left( {{Q^{ - 1}}\left( {{T_1}} \right)} \right)}^2}}}{{\alpha _B^2{{\left( {{Q^{ - 1}}\left( {{T_2}} \right)} \right)}^2} + \alpha _E^2{{\left( {{Q^{ - 1}}\left( {{T_1}} \right)} \right)}^2}}}.
 \end{align}
 Based on this problem, if a small value is assigned for the threshold of the legitimate receiver, the probability of outage is increased, and it is decreased for large values of the threshold of legitimate receiver. 
 
 \subsection{QPSK signaling and the other type of modulations}
 Quadrature phase shift keying (QPSK) is one of the most practical modulations that is used in digital communication systems. Symbol error probability (SEP) of the legitimate receiver for QPSK can be calculated as follows [31]:
   \begin{align}
&SE{P_B} = 1 - {\left[ {1 - Q\left( {\sqrt {\frac{{\alpha _B^2{{\left| {{h_B}} \right|}^2}P}}{{2{\sigma ^2}}}} } \right)} \right]^2} =\\& 2Q\left( {\sqrt {\frac{{\alpha _B^2{{\left| {{h_B}} \right|}^2}P}}{{2{\sigma ^2}}}} } \right) - {\left( {Q\left( {\sqrt {\frac{{\alpha _B^2{{\left| {{h_B}} \right|}^2}P}}{{2{\sigma ^2}}}} } \right)} \right)^2}. \nonumber
 \end{align}
 Similarly, for the adversary user, the SEP can be obtained as follows [31]:
    \begin{align}
SE{P_E} = 2Q\left( {\sqrt {\frac{{\alpha _E^2{{\left| {{h_E}} \right|}^2}P}}{{2{\sigma ^2}}}} } \right) - {\left( {Q\left( {\sqrt {\frac{{\alpha _E^2{{\left| {{h_E}} \right|}^2}P}}{{2{\sigma ^2}}}} } \right)} \right)^2}.
 \end{align}
 If Gray coding is used for bit mapping of symbols in QPSK signaling, the relation of BEP and SEP is as follows [31]:
 \begin{align}
       BEP \approx \frac{1}{2}SEP.
 \end{align}
 Based on this BEP value, the optimization problem for the QPSK signaling can be presented as follows:
  \begin{align}
 \begin{array}{l}
 \mathop {\min }\limits_P P\\
 _{}s.t. \,\, Q\left( {\sqrt {\frac{{\alpha _B^2{{\left| {{h_B}} \right|}^2}P}}{{2{\sigma ^2}}}} } \right) - \frac{1}{2}{\left( {Q\left( {\sqrt {\frac{{\alpha _B^2{{\left| {{h_B}} \right|}^2}P}}{{2{\sigma ^2}}}} } \right)} \right)^2} \le {T_1},\\\hspace{.52cm}
 Q\left( {\sqrt {\frac{{\alpha _E^2{{\left| {{h_E}} \right|}^2}P}}{{2{\sigma ^2}}}} } \right) - \frac{1}{2}{\left( {Q\left( {\sqrt {\frac{{\alpha _E^2{{\left| {{h_E}} \right|}^2}P}}{{2{\sigma ^2}}}} } \right)} \right)^2} \ge {T_2},
 \end{array}
 \end{align}
 In this problem, the threshold values for BEP are like the thresholds in problem (9). Since the error probability of the users should be a small value, we can eliminate the second term of the error probability relation. Therefore, the optimization problem for QPSK signaling can be approximated as follows:
   \begin{align}
 \begin{array}{l}
 \mathop {\min }\limits_P P\\
 _{}s.t. \,\, Q\left( {\sqrt {\frac{{\alpha _B^2{{\left| {{h_B}} \right|}^2}P}}{{2{\sigma ^2}}}} } \right) \le {T_1},\\\hspace{.52cm}
 Q\left( {\sqrt {\frac{{\alpha _E^2{{\left| {{h_E}} \right|}^2}P}}{{2{\sigma ^2}}}} } \right) \ge {T_2}.
 \end{array}
 \end{align}
 The feasibility constraint of this problem is similar to the BPSK case in problem (12), and for a feasible problem, the approximated optimum solution for the transmit power can be calculated as follows:
    \begin{align}
P_{opt}^{App} = \frac{{2{\sigma ^2}}}{{\alpha _B^2{{\left| {{h_B}} \right|}^2}}}{\left( {{Q^{ - 1}}\left( {{T_1}} \right)} \right)^2}.
 \end{align}
 In general case, we can expand our proposed solution for M-ary modulations like quadrature amplitude modulation (QAM). These modulation types have complicated bit error rate formula. Because of the complicated formulas, we cannot use these formulas in the optimization problem. The proposed solution for these problems is using boundary of the error probability. 
 One of the useful boundaries that is used for error probability is using minimum distance [31]. Minimum distance is the minimum Euclidean distance between symbols of a QAM signaling. Based on the minimum distance, the boundary for SEP can be obtained as follows:
    \begin{align}
SEP \le Q\left( {\frac{{{d_{\min }}}}{{2\sigma }}} \right),
\end{align} 
 where ${d_{\min }}$ is the minimum Euclidean distance between symbols. For this case, the optimization problem and its solution are like the BPSK case.
 Also Q function in the optimization problems does not have a closed-form formula. We can use the Chernoff bound as an approximation for Q function as follows [31]:
 \begin{align}
Q(x) \approx \frac{1}{2}\exp \left( { - \frac{{{x^2}}}{2}} \right),
 \end{align} 
where $\exp \left( . \right)$  is the exponential function. Based on this approximation, the approximated optimization problem for BPSK signaling as an example can be obtained as follows: 
    \begin{align}
 \begin{array}{l}
 \mathop {\min }\limits_P P\\
 _{}s.t. \,\, \alpha _B^2{\left| {{h_B}} \right|^2}P \ge  - 2{\sigma ^2}\ln \left( {2{T_1}} \right),\\\hspace{.58cm}
 \alpha _E^2{\left| {{h_E}} \right|^2}P \le  - 2{\sigma ^2}\ln \left( {2{T_2}} \right).
 \end{array}
 \end{align}
 For a feasible solution of this optimization problem, the approximated optimum transmit power value can be obtained as follows:
     \begin{align}
P_{opt}^{App} = \frac{{ - 2{\sigma ^2}\ln \left( {2{T_1}} \right)}}{{\alpha _B^2{{\left| {{h_B}} \right|}^2}}}.
 \end{align}
 
 \subsection{Multiple adversary users’ case}
 In this subsection, we assume that there are E adversary users in the network. The secrecy rate in this case is the minimum value with considering all of these adversary users. As mentioned in the previous section, the important challenge is the work point of the transmit power. In optimization problem for the multiple adversary users’ scenario, the BEP for all adversary users should be more than the threshold value. Therefore, the optimization problem for BPSK signaling can be presented as follows:
     \begin{align}
 \begin{array}{l}
 \mathop {\min }\limits_P P\\
 _{}s.t. \,\, Q\left( {\sqrt {\frac{{\alpha _B^2{{\left| {{h_B}} \right|}^2}P}}{{{\sigma ^2}}}} } \right) \le T_1,\\\hspace{.58cm}
 Q\left( {\sqrt {\frac{{\alpha _e^2{{\left| {{h_e}} \right|}^2}P}}{{{\sigma ^2}}}} } \right) \ge {T_2},\,\,e = 1,...,E,
 \end{array}
 \end{align}
where $h_e$ and $\alpha_e$ and   are the channel coefficient and path loss of adversary user $e$. For a feasible problem, the optimum solution is similar to the problem (11), and it is obtained in (13). 
According to the constraints of this problem, the feasibility condition for this problem can be presented as follows:
\begin{align}
\mathop {\max }\limits_e \left( {{{\left( {{Q^{ - 1}}\left( {{T_1}} \right)} \right)}^2} \times \alpha _e^2{{\left| {{h_e}} \right|}^2}} \right) \le {\left( {{Q^{ - 1}}\left( {{T_2}} \right)} \right)^2}\alpha _B^2{\left| {{h_B}} \right|^2}.
\end{align}
According to this constraint, the outage probability can be obtained as follows:
\begin{align}
\begin{array}{l}
{p_{Outage}} = 1 - \Pr \left\{ {\mathop {\max }\limits_e \left( {{{\left( {{Q^{ - 1}}\left( {{T_1}} \right)} \right)}^2} \times \alpha _e^2{{\left| {{h_e}} \right|}^2}} \right) \le } \right.\\
\,\,\,\,\,\,\,\,\,\,\,\,\,\,\,\,\,\,\left. {{{\left( {{Q^{ - 1}}\left( {{T_2}} \right)} \right)}^2}\alpha _B^2{{\left| {{h_B}} \right|}^2}} \right\}.
\end{array}
\end{align}
We define parameter $z$ as $z = \mathop {\max }\limits_e \left( {{{\left( {{Q^{ - 1}}\left( {{T_1}} \right)} \right)}^2}\alpha _e^2{{\left| {{h_e}} \right|}^2}} \right)$. To obtain this probability value, at first, the probability distribution function (PDF) of parameter $z$ should be obtained. It is the maximum value of exponential random variables and based on probability theories in [32], the PDF can be obtained as follows:
\begin{align}
&{f_Z}\left( {z = \mathop {\max }\limits_e \left( {{{\left( {{Q^{ - 1}}\left( {{T_1}} \right)} \right)}^2}\alpha _e^2{{\left| {{h_e}} \right|}^2}} \right)} \right) =\\& \sum\limits_{e = 1}^E {\left( {{{F'}_e}(z)\prod\limits_{i \ne e} {{F_e}(z)} } \right)},\nonumber
\end{align}
where ${F_e}(.)$ is the exponential cumulative distribution function of ${\left( {{Q^{ - 1}}\left( {{T_1}} \right)} \right)^2}\alpha _e^2{\left| {{h_e}} \right|^2}$
with parameter ${\lambda _e} = 1/\left( {{{\left( {{Q^{ - 1}}\left( {{T_1}} \right)} \right)}^2}\alpha _e^2} \right)$ and ${F'_e}(.)$ is the first order derivative function of ${F_e}(.)$. Therefore, the outage probability can be rewritten as follows:
\begin{align}
{p_{Outage}} = 1 - \Pr \left\{ {z \le y} \right\} = 1 - \int\limits_0^{ + \infty } {\int\limits_0^y {{f_Z}(z){f_Y}(y)dzdy} },
\end{align}
where $y = {\left( {{Q^{ - 1}}\left( {{T_2}} \right)} \right)^2}\alpha _B^2{\left| {{h_B}} \right|^2}$ and ${f_Y}(y)$ is the exponential PDF with parameter ${\lambda _y} = 1/\left( {{{\left( {{Q^{ - 1}}\left( {{T_2}} \right)} \right)}^2}\alpha _B^2} \right)$.
By substituting the distribution functions in (30), the outage probability can be obtained as (31), in the top of next page.
\begin{figure*}[t]    	
	\begin{align}
\begin{array}{l}
{p_{Outage}} = 1 - \int\limits_0^{ + \infty } {\int\limits_0^y {\left( {\sum\limits_{e = 1}^E {\left( {{\lambda _e}{e^{ - {\lambda _e}z}}\prod\limits_{i \ne e} {\left( {1 - {e^{ - {\lambda _e}z}}} \right)} } \right)} } \right)\left( {{\lambda _y}{e^{ - {\lambda _y}y}}} \right)dzdy} } \\
= 1 - \int\limits_0^{ + \infty } {\left. {\prod\limits_e {\left( {1 - {e^{ - {\lambda _e}z}}} \right)} } \right|_0^y\left( {{\lambda _y}{e^{ - {\lambda _y}y}}} \right)dy}  = \int\limits_0^{ + \infty } {\prod\limits_e {\left( {1 - {e^{ - {\lambda _e}y}}} \right)} \left( {{\lambda _y}{e^{ - {\lambda _y}y}}} \right)dy} \\
= 1 - \int\limits_0^{ + \infty } {\left( {1 - {e^{ - {\lambda _1}y}}} \right)\left( {1 - {e^{ - {\lambda _2}y}}} \right)...\left( {1 - {e^{ - {\lambda _E}y}}} \right)\left( {{\lambda _y}{e^{ - {\lambda _y}y}}} \right)dy} \\
= 1 - \int\limits_0^{ + \infty } {\left( {{\lambda _y}{e^{ - {\lambda _y}y}} - {\lambda _y}{e^{ - ({\lambda _1} + {\lambda _y})y}} - {\lambda _y}{e^{ - ({\lambda _2} + {\lambda _y})y}} - ... + {\lambda _y}{e^{ - ({\lambda _1} + {\lambda _2} + {\lambda _y})y}} + ... + {{( - 1)}^E}{\lambda _y}{e^{ - ({\lambda _1} + ... + {\lambda _E} + {\lambda _y})y}}} \right)dy} \\
= \frac{{{\lambda _y}}}{{{\lambda _1} + {\lambda _y}}} + \frac{{{\lambda _y}}}{{{\lambda _2} + {\lambda _y}}} + ... + \frac{{{\lambda _y}}}{{{\lambda _E} + {\lambda _y}}} - \frac{{{\lambda _y}}}{{{\lambda _1} + {\lambda _2} + {\lambda _y}}} - ...\\
= \sum\limits_{i = 1}^E {\frac{{{\lambda _y}}}{{{\lambda _i} + {\lambda _y}}}}  - \sum\limits_{\scriptstyle i,j\hfill\atop
	\scriptstyle i \ne j\hfill}^{} {\frac{{{\lambda _y}}}{{{\lambda _i} + {\lambda _j} + {\lambda _y}}}}  + \sum\limits_{\scriptstyle i,j,k\hfill\atop
	\scriptstyle i \ne j \ne k\hfill}^{} {\frac{{{\lambda _y}}}{{{\lambda _i} + {\lambda _j} + {\lambda _k} + {\lambda _y}}}}  - ... - {( - 1)^E}\frac{{{\lambda _y}}}{{{\lambda _1} + {\lambda _2} + ... + {\lambda _E} + {\lambda _y}}}.
\end{array}
	\end{align}
	\hrule
\end{figure*}
\subsection{Adversary users with unknown mode}
At the previous subsections, we assume that the adversary user is a passive user and he/she just eavesdrops in the channel of the legitimate receivers. Generally, the adversary users can be passive or active. In the active mode, the adversary users can send jamming signal to disrupt in the communication of the legitimate users. 
In the most of the works about physical layer security, the modes of the adversary users are known and the system model and power allocation are obtained based on it. In this subsection, we assume that the modes of the adversary users are unknown, and there is no more information about the adversary users. For this scenario, we assume the worst case for the network. Because the modes of the adversary users are unknown, we assume the best case for the adversary users, and the result is the worst case for the network.
We consider the interference value of the adversary users like additive Gaussian noise signal. Also, it is assumed that the adversary users can cancel this interference value, because there is no more information about them and also, this assumption is the worst case for the network.
The best case for the adversary users is assuming the best user as a passive user and the others as the active users [25]. The best adversary user has the largest channel gain for eavesdropping. In this case, the optimization problem can be presented as follows:
     \begin{align}
\begin{array}{l}
\mathop {\min }\limits_P P\\
_{}s.t. \,\, Q\left( {\sqrt {\frac{{\alpha _B^2{{\left| {{h_B}} \right|}^2}P}}{{{\sigma ^2} + I}}} } \right) \le {T_1},\\\hspace{.58cm}
Q\left( {\sqrt {\frac{{\mathop {\max }\limits_e \left\{ {\alpha _e^2{{\left| {{h_e}} \right|}^2}} \right\}P}}{{{\sigma ^2}}}} } \right) \ge {T_2},
\end{array}
\end{align}
where $I$ is the interference value from the active adversary users, and it is assumed that it is a Gaussian random variable. In this problem, we assume that the adversary users want to receive data information with high performance. The adversary user with high performance is a passive user and the others send jamming signal to disrupt in the communication of the legitimate receivers. Therefore, the other adversary users do not eavesdrop and they send jamming signals to make interference at the legitimate receivers. 

Based on the previous subsection, for a feasible solution, the optimum transmit power can be obtained as follows:
     \begin{align}
{P_{opt}} = \frac{{{\sigma ^2} + I}}{{\alpha _B^2{{\left| {{h_B}} \right|}^2}}}{\left( {{Q^{ - 1}}\left( {{T_1}} \right)} \right)^2}.
\end{align}
As we can see in this problem, the optimum transmit power is increased with interference, it is needed to use more transmit power to reach the proper performance in the presence of the interference of the adversary users. 
The feasibility constraint for this problem can be presented as follows:
     \begin{align}
&\left( {{\sigma ^2} + I} \right){\left( {{Q^{ - 1}}\left( {{T_1}} \right)} \right)^2} \times \mathop {\max }\limits_e \left( {\alpha _e^2{{\left| {{h_e}} \right|}^2}} \right) \le\\&\hspace{1.5cm}{\sigma ^2}{\left( {{Q^{ - 1}}\left( {{T_2}} \right)} \right)^2}\alpha _B^2{\left| {{h_B}} \right|^2}.\nonumber
\end{align}
Based on this constraint, the outage probability is like (31) with the following parameters:
     \begin{align}
\left\{ \begin{array}{l}
{\lambda _y} = 1/\left( {{\sigma ^2}{{\left( {{Q^{ - 1}}\left( {{T_2}} \right)} \right)}^2}\alpha _B^2} \right),\\
{\lambda _e} = 1/\left( {\left( {{\sigma ^2} + I} \right){{\left( {{Q^{ - 1}}\left( {{T_1}} \right)} \right)}^2}\alpha _e^2} \right).
\end{array} \right.
\end{align}
\subsection{Cooperative adversary users}
The adversary users can have cooperation with each other to reach a better performance. The adversary users can use the techniques that are used in wireless communication systems like maximum ratio combining (MRC). We assume that they have a central node to receive and process all data of the adversary users.  If they use the MRC technique, the BEP constraint in the optimization problem for the adversary users can be presented as follows:
     \begin{align}
Q\left( {\sqrt {\frac{{\sum\limits_e {\left\{ {\alpha _e^2{{\left| {{h_e}} \right|}^2}} \right\}P} }}{{{\sigma ^2}}}} } \right) \ge {T_2}.
\end{align}
Based on this constraint, the optimum transmit power value for a feasible problem is similar to (13).
In this case, the outage probability can be obtained as follows:
     \begin{align}
\begin{array}{l}
{p_{Outage}} = 1 - \Pr \left\{ {{{\left( {{Q^{ - 1}}\left( {{T_1}} \right)} \right)}^2} \times \sum\limits_e {\left( {\alpha _e^2{{\left| {{h_e}} \right|}^2}} \right)}  \le } \right.\\
\left. {{{\left( {{Q^{ - 1}}\left( {{T_2}} \right)} \right)}^2}\alpha _B^2{{\left| {{h_B}} \right|}^2}} \right\}
\end{array}
\end{align}
To obtain the outage probability in a closed-form, we define $w = {\left( {{Q^{ - 1}}\left( {{T_1}} \right)} \right)^2} \times \sum\limits_e {\left( {\alpha _e^2{{\left| {{h_e}} \right|}^2}} \right)} $, the PDF of variable $w$  should be calculated. It is summation of $E$ exponential random variables. Based on the theorems of probability random variables [32], it can be obtained as follows:
     \begin{align}
&{f_W}\left( {w = {{\left( {{Q^{ - 1}}\left( {{T_1}} \right)} \right)}^2} \times \sum\limits_e {\left( {\alpha _e^2{{\left| {{h_e}} \right|}^2}} \right)} } \right) =\\& \left( {\prod\limits_e {{\lambda _e}} } \right)\sum\limits_{e = 1}^E {\frac{{{e^{ - {\lambda _e}w}}}}{{\prod\limits_{i,i \ne e} {\left( {{\lambda _i} - {\lambda _e}} \right)} }}} ,\nonumber
\end{align}
where ${\lambda _e} = 1/\left( {{{\left( {{Q^{ - 1}}\left( {{T_1}} \right)} \right)}^2}\alpha _e^2} \right)$ is the parameter of the exponential random variable  ${\left( {{Q^{ - 1}}\left( {{T_1}} \right)} \right)^2}\left( {\alpha _e^2{{\left| {{h_e}} \right|}^2}} \right).$ Similar to (30), the outage probability can be rewritten as follows:
     \begin{align}
\begin{array}{l}
{p_{Outage}} = 1 - \Pr \left\{ {w \le y} \right\} = 1 - \int\limits_0^{ + \infty } {\int\limits_0^y {{f_W}(w){f_Y}(y)dwdy} }  \\= 1 - \int\limits_0^{ + \infty } {\int\limits_0^y {\left( {\prod\limits_e {{\lambda _e}} } \right)\sum\limits_{e = 1}^E {\frac{{{e^{ - {\lambda _e}w}}}}{{\prod\limits_{i,i \ne e} {\left( {{\lambda _i} - {\lambda _e}} \right)} }}} \left( {{\lambda _y}{e^{ - {\lambda _y}y}}} \right)dwdy} } \\
= 1 - \int\limits_0^{ + \infty } {\left( {\prod\limits_e {{\lambda _e}} } \right)\sum\limits_{e = 1}^E {\frac{1}{{{\lambda _e}}}\frac{{\left( {1 - {e^{ - {\lambda _e}y}}} \right)}}{{\prod\limits_{i,i \ne e} {\left( {{\lambda _i} - {\lambda _e}} \right)} }}} \left( {{\lambda _y}{e^{ - {\lambda _y}y}}} \right)dy}  =\\ 1 - \left( {\prod\limits_e {{\lambda _e}} } \right)\sum\limits_{e = 1}^E {\frac{1}{{\prod\limits_{i,i \ne e} {\left( {{\lambda _i} - {\lambda _e}} \right)} }}\left( {\frac{1}{{{\lambda _e} + {\lambda _y}}}} \right)} .
\end{array}
\end{align}
Based on this equation, for two and three adversary users, the outage probability can be obtained as follows:

For $E=2$:
     \begin{align}
\begin{array}{l}
{p_{Outage}} = 1 - \left( {{\lambda _1}{\lambda _2}} \right)\left( {\frac{1}{{{\lambda _2} - {\lambda _1}}} \times \frac{1}{{{\lambda _1} + {\lambda _y}}} + \frac{1}{{{\lambda _1} - {\lambda _2}}} \times \frac{1}{{{\lambda _2} + {\lambda _y}}}} \right)\\
= 1 - \left( {\frac{{{\lambda _1}{\lambda _2}}}{{{\lambda _2} - {\lambda _1}}}} \right)\left( {\frac{1}{{{\lambda _1} + {\lambda _y}}} - \frac{1}{{{\lambda _2} + {\lambda _y}}}} \right).
\end{array}
\end{align}

For $E=3$:
\begin{align}
\begin{array}{*{20}{l}}
{{p_{Outage}} = 1 - \left( {{\lambda _1}{\lambda _2}{\lambda _3}} \right)\left( {\frac{1}{{\left( {{\lambda _2} - {\lambda _1}} \right)\left( {{\lambda _3} - {\lambda _1}} \right)}} \times \frac{1}{{{\lambda _1} + {\lambda _y}}} + } \right.}\\
{\left. {\frac{1}{{\left( {{\lambda _1} - {\lambda _2}} \right)\left( {{\lambda _3} - {\lambda _2}} \right)}} \times \frac{1}{{{\lambda _2} + {\lambda _y}}} + \frac{1}{{\left( {{\lambda _1} - {\lambda _3}} \right)\left( {{\lambda _2} - {\lambda _3}} \right)}} \times \frac{1}{{{\lambda _3} + {\lambda _y}}}} \right).}
\end{array}
\end{align}

\section{Improving outage probability by using multiple antennas transmitter and channel coding}\label{multiple antennas}
In this subsection, two approaches are proposed to improve the outage probability. By using the multiple antenna transmitter, it is possible to apply beamforming to send artificial noise and data signals simultaneously and degrade the performance of the adversary users. Also, channel coding techniques are adopted to improve BEP in digital communication systems. In the proposed system, we can use channel coding to reduce BEP of the legitimate receiver and improve the outage probability. 
\subsection{Multiple antenna transmitter and sending artificial noise}
One of the most useful techniques in physical layer security is using beamforming and sending artificial noise to confuse the adversary users. In this section, we assume that the transmitter is equipped with M antennas to send data and artificial noise signals via beamforming. In this system, the received signals at the legitimate and adversary users can be obtained as follows:
\begin{align}
&{y_B} = {\bf{h}}_B^T{{\bf{w}}_d}{x_d} + {\bf{h}}_B^T{{\bf{w}}_{an}}{x_{an}} + {n_B},\\& {y_e} = {\bf{h}}_e^T{{\bf{w}}_d}{x_d} + {\bf{h}}_e^T{{\bf{w}}_{an}}{x_{an}} + {n_e}, \, e = 1,...,E,
\end{align}
where ${{\bf{h}}_B}$ and ${{\bf{h}}_e}$ are $M\times 1$ channel vectors from the transmitter to the legitimate user and adversary user $e$, ${{\bf{w}}_d}$ and ${{\bf{w}}_{an}}$ are beamforming vectors for data and artificial noise signals, $x_d$ and $x_{an}$ are data and artificial noise signals with power values ${P_d}$ and ${P_{an}}$, and ${n_{B}}$ and ${n_{e}}$ are the zero-mean complex Gaussian noise signals at legitimate user and adversary user $e$ with variances ${\sigma ^2}$.

The beamforming vectors are designed by using the channel vectors. The conventional beamforming is applied at the transmitter for data signal, and artificial noise is sent in the null space of the legitimate user. Then, the beamforming vectors are calculated as follows:
\begin{align}
&{\bf{h}}_B^H{{\bf{w}}_d} = \left\| {{{\bf{h}}_B}} \right\|,\\& {\bf{h}}_B^H{{\bf{w}}_{an}} = 0, \,\, M > 1.
\end{align}
where $\left\| . \right\|$ is the norm function. Channel coefficients are assumed i.i.d. zero-mean circularly symmetric complex Gaussian random variables.
Based on the beamforming vectors and the defined threshold values in the previous section, the optimization problem for this scenario can be proposed as follows:

     \begin{align}
\begin{array}{l}
\mathop {\min }\limits_{{P_d},{P_{an}}} \left( {{P_d} + {P_{an}}} \right)\\
s.t. \,\, Q\left( {\sqrt {\frac{{{{\left\| {{{\bf{h}}_B}} \right\|}^2}{P_d}}}{{{\sigma ^2}}}} } \right) \le {T_1},\\ \hspace{.58cm}
Q\left( {\sqrt {\frac{{{{\left\| {{\bf{h}}_e^H{{\bf{w}}_d}} \right\|}^2}{P_d}}}{{{\sigma ^2} + {{\left\| {{\bf{h}}_e^H{{\bf{w}}_{an}}} \right\|}^2}{P_{an}}}}} } \right) \ge {T_2}, \,\, e = 1,...,E.
\end{array}
\end{align}
This optimization problem can be rewritten as follows:
     \begin{align}
\begin{array}{l}
\mathop {\min }\limits_{{P_d},{P_{an}}} \left( {{P_d} + {P_{an}}} \right)\\
s.t. \,\, {\sigma ^2}{\left( {{Q^{ - 1}}\left( {{T_1}} \right)} \right)^2} - {\left\| {{{\bf{h}}_B}} \right\|^2}{P_d} \le 0,\\ \hspace{.58cm}
{\left\| {{\bf{h}}_e^H{{\bf{w}}_d}} \right\|^2}{P_d} - {\left( {{Q^{ - 1}}\left( {{T_2}} \right)} \right)^2}\times\\\hspace{.99cm}\left( {{\sigma ^2} + {{\left\| {{\bf{h}}_e^H{{\bf{w}}_{an}}} \right\|}^2}{P_{an}}} \right) \le 0, \,\, e = 1,...,E.
\end{array}
\end{align}
The above optimization problem is a convex problem and it can be solved by using CVX MATLAB toolbox or the other methods such as the dual approach [33].
\subsection{ Channel coding}
For some services like signaling, low value for BEP is needed for the legitimate users. Based on the previous sections, low threshold value for BEP of the legitimate users increases the outage probability. High outage probability value for services with high priority is not acceptable in a real network. Therefore, to decrease the outage probability of the network, we propose to apply channel coding techniques.  
In this paper, we assume that the users of the network are not equipped with high performance processors and they are not high power consumption. Therefore, in this section, we use low complexity channel coding scheme to enhance the performance of the network.
The block based codes are used for these services. Channel coding design is based on the threshold value for BEP of the adversary users. 
 In this paper, without loss of generality we assume that the error correction is used at the receivers. The block code that is used in this paper is assumed with length $N$ and it can correct $t$ errors in each block. If  $t$ or less than  $t$ errors are occurred in each block, this channel coding technique can correct these errors but if the number of errors is more than $t$, this channel coding technique cannot correct errors in each block. Therefore, based on the threshold values for legitimate and adversary users, the proposed values for length $N$ and $t$ for a block coding technique can be obtained as follows:
     \begin{align}
\left\{ \begin{array}{l}
BE{P_e} \times N >  > t,\\
BE{P_B} \times N <  < t.
\end{array} \right.
\end{align}
By using these inequalities, the legitimate user has less than $t$ errors in each block with high probability, and the adversary users have more than $t$ errors in each block with high probability. Therefore, the block code with capability of correction of $t$ errors can correct the errors of the legitimate users in each block but it cannot correct the errors of the adversary users. Also, in this scenario there is no limitation for decoding technique at the adversary users and it is assumed that the adversary users know about coding and decoding techniques exactly. 

\section{ Simulation results}\label{results}
Based on the system model, we assume that there are one transmitter and legitimate receiver and some adversary users in the network. We assume that the channel coefficients are i.i.d. zero-mean circularly symmetric complex Gaussian random variables with unit variance. All simulation parameters are presented in Table 1.
  \begin{table}[t] 
  	\centering
	\caption{Simulation parameters}
	\begin{tabular}{|c |c |} 
		\hline
		${\sigma ^2}$ &   $0.01$ \\ [0.5ex]  
		\hline
		$T_1$  &  It is variable \\ 
		\hline
		$T_2$ &  $0.05$ \\
		\hline
		$E$&  $1$, $2$ and $3$\\
		\hline
		$M$&   $4$  \\
		\hline
		$N$& $63$\\
		\hline
		$t$& $1$\\
		\hline
	\end{tabular}
	\label{table}
\end{table}

\begin{figure}[h]
	\includegraphics[width=3.2in,height=2.5in]{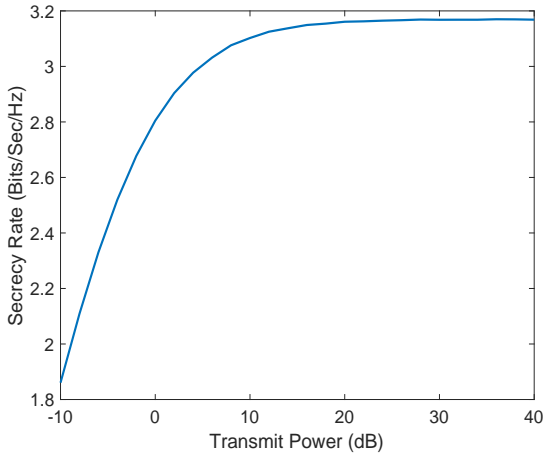}% \vspace{3cm}
	\hspace{-2.8cm}		\caption{The secrecy rate \textit{vs}. Transmit power.}
	\label{1}
\end{figure}
\begin{figure}[h]
	\includegraphics[width=3.2in,height=2.5in]{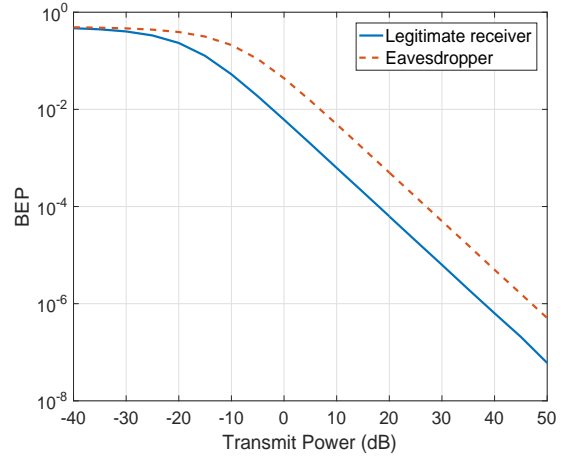}% \vspace{3cm}
	\hspace{-2.8cm}		\caption{BEP \textit{vs}. Transmit power.}
	\label{2}
\end{figure}
In Fig. 1, the secrecy rate that is presented in (1) is shown for different values of the transmit power. As shown in this figure, the secrecy rate increases with increasing the transmit power, but it is saturated for the large values of the transmit power, i.e., the secrecy rate is converged to the maximum value.

In Fig. 2, BEP of the legitimate receiver and the adversary user are presented. For the large values of the transmit power, the BEPs for the legitimate receiver and the adversary user have small values and they can be acceptable for some networks or services. Therefore, increasing transmit power to the large values is not recommended in these systems, and it is needed to have some constraints on BEP of the users. 

\begin{figure}[h]
	\includegraphics[width=3.2in,height=2.5in]{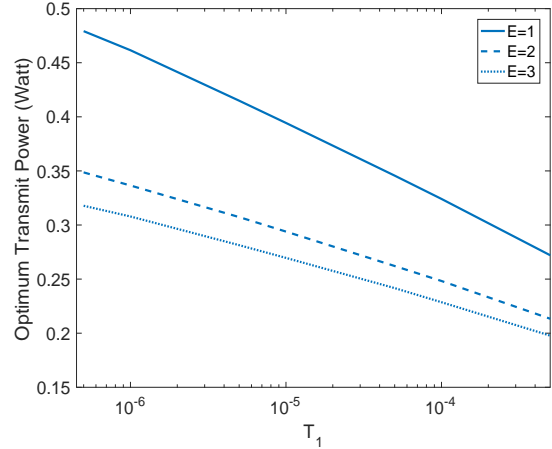}% \vspace{3cm}
	\hspace{-2.8cm}		\caption{The optimum power value for different number of the adversary users.}
	\label{3}
\end{figure}
 \begin{figure}[h]
	\includegraphics[width=3.2in,height=2.5in]{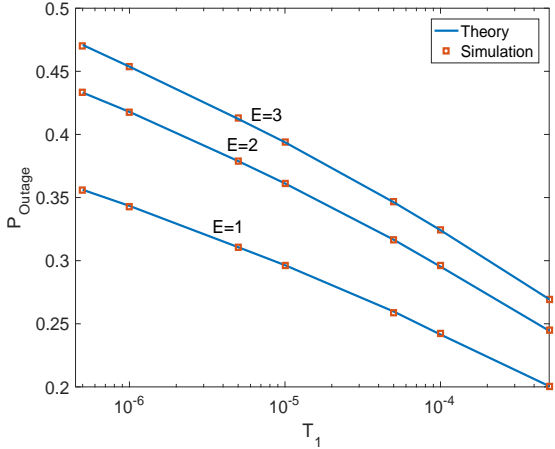}% \vspace{3cm}
	\hspace{-2.8cm}		\caption{The outage probability for different number of the adversary users.}
	\label{4}
\end{figure}
The optimum transmit power for different values of $T_1$ is presented in Fig. 3. If the threshold value for the error probability of the legitimate user increases, the optimum power value decreases, because the bit error threshold value for the legitimate user can be satisfied by a lower transmit power value. Also, the other important result of this figure is decreasing the power value for multi adversary users’ case. If there are more adversary users in the network, the ability that they can reach the desired error probability is increased. Therefore, the transmit power value for the multi adversary users’ case is decreased to prevent decreasing the error probability of adversary users. On the other hand, based on the optimum transmit power formula, it is not depended on the number of adversary users. The point is the feasibility of the optimization problem for the multi adversary users’ case. In this case, the optimization problem is feasible for the large values of the channel power gain for the legitimate user. For the large channel power gains of the legitimate receiver, the optimum transmit power has smaller values.

 The outage probability is shown in Fig. 4. The outage probability increases for the multi adversary users’ case and decreases for large bit error probability threshold for the legitimate receiver. Moreover, the simulation results confirm the theoretical formulas for the outage probability for the different number of the adversary users. 
 
 \begin{figure}[h]
 	\includegraphics[width=3.2in,height=2.5in]{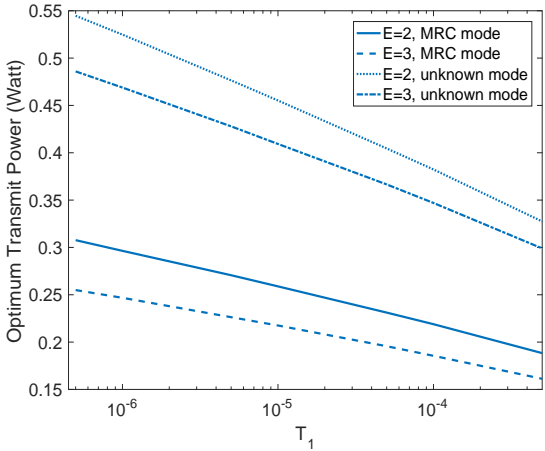}% \vspace{3cm}
 	\hspace{-2.8cm}		\caption{The optimum power value for cooperative adversary users and unknown mode.}
 	\label{5}
 \end{figure}

 \begin{figure}[h]
	\includegraphics[width=3.2in,height=2.5in]{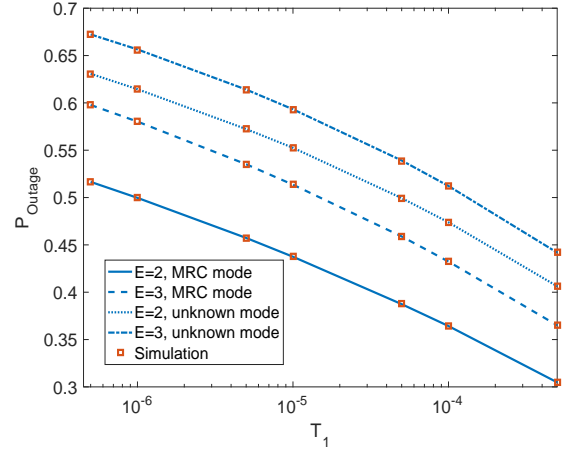}% \vspace{3cm}
	\hspace{-2.8cm}		\caption{The outage probability for cooperative adversary users and unknown mode.}
	\label{6}
\end{figure}

 \begin{figure}[h]
	\includegraphics[width=3.2in,height=2.5in]{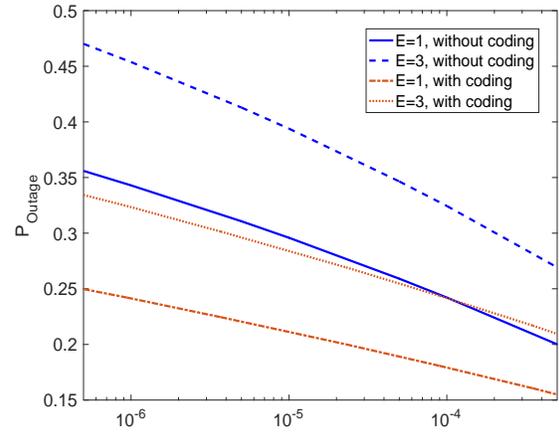}% \vspace{3cm}
	\hspace{-2.8cm}		\caption{Outage probability with channel coding.}
	\label{7}
\end{figure}

 In Fig. 5, the optimum transmit power is presented for the cooperative adversary users and unknown mode. As we can see, optimum power for the MRC mode is less than optimum power for the unknown mode. In MRC, the adversary users combine the received signals to reach a better performance, therefore, the optimum transmit power should be a small value to prevent signal detection of the adversary users with desired performance.

 In Fig. 6, the outage probability is illustrated for the cooperative adversary users and unknown modes. The outage probability for the cooperative adversary users’ case is less than that of the adversary users with unknown modes. In the adversary users with the unknown modes, we assume that they send jamming signal to increase the BEP of the legitimate users, and in this case we assume the worst case for the network. Therefore, the outage probability in this case has a larger value. Also, the simulations confirm the theoretical formulas for the outage probability for the different scenarios.

 In Fig. 7, the outage probability is shown with channel coding technique. The Hamming code is used for this scenario. The threshold value for the adversary users is assumed 0.05. Based on this threshold value, the parameters of Hamming code are selected as $N=63$  and $t=1$. As we can see in Fig. 7, the outage probability is decreased for the different number of the adversary users. 

\begin{figure}[h]
	\includegraphics[width=3.2in,height=2.5in]{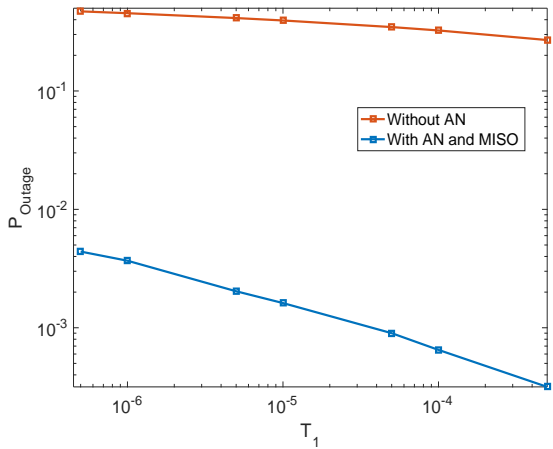}% \vspace{3cm}
	\hspace{-2.8cm}		\caption{Outage probability with using beamforming and sending artificial noise.}
	\label{8}
\end{figure}

In Fig. 8, the outage probability is shown for multiple antenna transmitter. As seen, when beamforming with artificial noise signal is used in the network the outage probability is decreased efficiently.

 \begin{figure}[h]
 	\includegraphics[width=3.2in,height=2.5in]{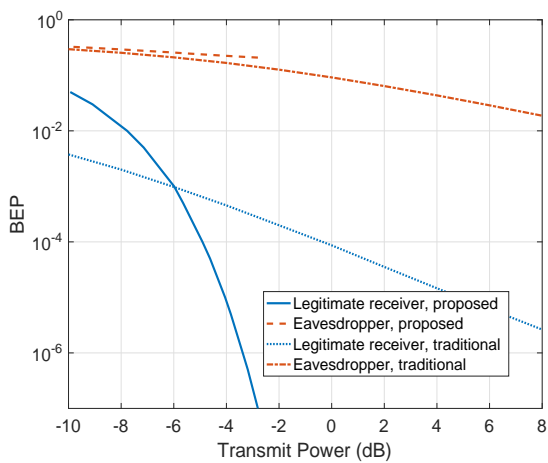}% \vspace{3cm}
 	\hspace{-2.8cm}		\caption{BEP of the legitimate and adversary users.}
 	\label{9}
 \end{figure}

 \begin{figure}[h]
 	\includegraphics[width=3.2in,height=2.5in]{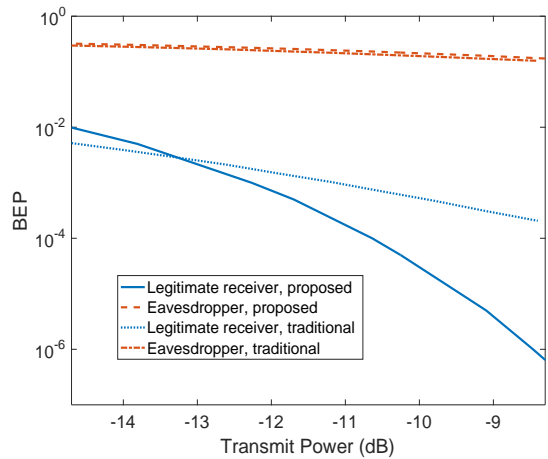}% \vspace{3cm}
 	\hspace{-2.8cm}		\caption{BEP of the legitimate and adversary users by using beamforming and artificial noise.}
 	\label{10}
 \end{figure}

 BEP for the legitimate and adversary users are shown in Figs. 9 and 10 for traditional and proposed methods. Fig. 9 shows BEP for the single antenna transmitter. Based on this figure, the proposed method has lower BEP for the legitimate user. Also, Fig. 10 shows BEP for multiple antenna transmitter with beamforming and sending artificial noise signal, and BEP of the legitimate user in the proposed method has lower value. The minimum secrecy rate for simulation of the traditional method in these figures is ${R_{min}} = 2$. As we can see in these figures, the proposed method needs more than 5dB lower power for different scenarios in high SNR values. 
 
 \section{Conclusion}\label{Conclusion}
 In this paper, based on BEP, a new power allocation approach was proposed for physical layer security systems. Threshold values were defined for BEP of the legitimate and adversary users and the power allocation problem and the optimum transmit power were obtained based on these threshold values. Moreover, the proposed approach was applied in the presence of the multi adversary users, cooperative adversary users and adversary users with unknown modes. In this paper, a new definition for the outage probability was defined and it was obtained analytically for different scenarios. A simple channel coding scheme and artificial noise were used to improve the outage probability of the network. The BEP for the proposed method is better than that of a traditional method that is used just the secrecy rate.

\end{document}